\documentclass[prl,twocolumn,showpacs,preprintnumbers,amsmath,amssymb]{revtex4}

\usepackage{graphicx}

\begin{document}

\title{Dark stationary matter-waves via parity-selective evaporation \\
in a Tonks-Girardeau gas}

\author{H. Buljan,$^{1}$ O. Manela,$^{2}$ R. Pezer,$^{1}$ 
A. Vardi,$^{3}$ and M. Segev$^{2}$}
\affiliation{$^{1}$Department of Physics, 
University of Zagreb, PP 332, Zagreb, Croatia}
\affiliation{$^{2}$Physics Department, Technion - 
Israel Institute of Technology, Haifa 32000, Israel}
\affiliation{$^3$Department of Chemistry, Ben Gurion 
University of Negev, Beer Sheva 84105, Israel}
\date{\today}

\begin{abstract}
We propose a scheme for observing dark stationary waves in a 
Tonks-Girardeau (TG) gas. 
The scheme is based on parity-selective dynamical "evaporation" 
of the gas via a time-dependent potential, which excites the 
gas from its ground state towards a desired specially-tailored 
many-body state. 
These excitations of the TG gas are analogous 
to linear partially coherent nondiffracting beams in optics, as evident from 
the mapping between the quantum dynamics of the TG gas and the propagation of 
incoherent light in one-dimensional linear photonic structures.
\end{abstract}

\maketitle

Trapped bose gases confined to one-dimensional (1D) geometry are highly 
attractive for studying quantum many-body dynamics 
because pertinent models yield exact solutions 
\cite{Girardeau1960,Lieb,Lenard1964,Girardeau2000}, these regimes are 
experimentally accessible \cite{OneD,Paredes2004,Kinoshita2004,Kinoshita2006}, 
and quantum effects are enhanced \cite{Olshanii,Petrov,Dunjko}. 
Tonks-Girardeau (TG) gas is a system of 1D bosons 
with "impenetrable core" repulsive interactions \cite{Girardeau1960}. 
In the fashion of the Pauli exclusion principle, 
"impenetrable cores" prevent bosons to occupy the 
same position in space, which causes TG gas to exhibit fermionic 
properties. This similarity is manifested in the mapping 
between 1D noninteracting fermions and TG bosons \cite{Girardeau1960,Girardeau2000}. 
From the properties of atomic interactions in tight atomic waveguides 
\cite{Olshanii} it follows that the TG regime can be reached 
at low temperatures, low linear densities or stronger effective 
interactions \cite{Olshanii,Petrov,Dunjko}. 
The experimental realizations of the TG gas in 2004 
\cite{Kinoshita2004,Paredes2004} boosted the physical 
relevance of the model. 
A recent experiment has demonstrated that such gas does 
not relax to the thermodynamic equilibrium even after numerous 
collisions \cite{Kinoshita2006}, due to the integrability of 
the underlying model. This work inspired a theoretical study of 1D 
"impenetrable core" bosons on a lattice, which suggested that 
the system can undergo irreversible relaxation to a steady state
carrying more memory of the initial conditions than the 
usual thermodynamic equilibrium \cite{Rigol2006}. 
Some aspects of the TG quantum dynamics have been studied theoretically 
within the context of so-called "dark solitons" \cite{Girardeau2000,Busch2003}, 
matter-wave interference \cite{Girardeau2000a}, 1D expansion \cite{Ohberg2002,
Rigol2005}, irregular dynamics \cite{Berman2004}, and coherent states 
\cite{Minguzzi2005}. In this work, we study the dynamical tailoring of 
the TG gas via a time-dependent potential to produce dark 
stationary states, as well as point out the relation between TG dynamics 
and the propagation of incoherent light in linear photonic structures.

The Fermi-Bose mapping \cite{Girardeau1960,Girardeau2000}, applicable both 
in the static \cite{Girardeau1960} and time-dependent case \cite{Girardeau2000}, 
prescribes the construction of the exact many-body wavefunction of the 
TG gas from single-particle (SP) wavefunctions, which obey a set of uncoupled 
linear SP Schr\" odinger equations. In Ref. \cite{Girardeau2000} Girardeau and 
Wright discuss the dynamics of the TG gas within the context of dark solitons 
\cite{Dum1998,ExpDark,Busch2000,Muryshev2002}. Dark solitons are fundamental nonlinear 
excitations, which have been studied theoretically within the nonlinear 
mean-field theories applicable for weakly interacting gases 
\cite{Dum1998,Busch2000,Muryshev2002}, and observed experimentally in these 
regimes \cite{ExpDark}. 
For a strongly interacting TG gas on a ring, Girardeau and Wright \cite{Girardeau2000} 
noticed that if the many-body wavefunction is constructed solely from the 
odd-parity SP eigenstates of the system, the SP density will have a dip 
at zero, similar in structure to dark-solitons. However, such a specially 
structured many-body state is unlikely to occur without deliberate 
preparation, since even and odd parity SP states of that system are 
intermingled when ordered with respect to energy (see the discussion in 
Ref. \cite{Girardeau2000}). 
In the study of Busch and Huyet \cite{Busch2003}, the collapses and reappearances 
of TG dark solitonlike structures in an harmonic trap are attributed to the 
mixture of the odd and the even components in the excitation. Generally, the 
SP eigenstates in parity-invariant 1D potentials can be chosen to be either 
even or odd, which makes them candidates for observing dark stationary 
structures in the TG gas. However, for their {\em experimental realization} 
under such confinement, it is essential to separate components of different parity.

Here we propose a scheme for observing dark stationary waves in a TG gas. 
A time-dependent potential is used to selectively "evaporate" the even 
component of the many-body wavefunction, thereby creating a dark stationary 
wave. Such excitation of the strongly interacting 
TG gas is in fact an excited {\em many-body eigenstate} of the system, 
which distinguishes it from dark-solitons of the nonlinear 
mean-field equations applicable for weak interactions \cite{Dum1998,ExpDark,Busch2000,Muryshev2002}. 
We point out that such excited eigenstates of the TG gas are analogous to 
linear partially coherent nondiffracting beams in optics \cite{Turunen,Shchegrov}, 
as evident from the mapping between the quantum dynamics of the TG gas and the 
propagation of incoherent light in one-dimensional linear photonic structures, 
presented in this Letter.

We consider $N$ impenetrable bosons, confined within a 1D external potential 
$V_{ext}(x,t)$. The fully symmetrized many-body wavefunction describing 
the system, $\psi_B(x_1,\ldots,x_N,t)$, is constructed according to 
the Fermi-Bose mapping \cite{Girardeau1960,Girardeau2000}. Let $\psi(\xi,\tau)$ 
denote a set of orthonormal SP wavefunctions obeying the set of uncoupled 
linear Schr\" odinger equations,

\begin{equation}
i\frac{\partial \psi_m}{\partial \tau}=
\left [ -\frac{\partial^2 }{\partial \xi^2}+
V(\xi,\tau) \right ] \psi_m(\xi,\tau), \ m=1,\ldots,N.
\label{master}
\end{equation}
In order to unify notation and discuss the equivalence with optics, 
we find it convenient to use dimensionless units. The boson mass $m$ 
and the (arbitrary) choice of spatial lengthscale $x_0$ ($\xi=x/x_0$) 
determine the units of time $t_0=2mx_0^2/\hbar$ ($\tau=t/t_0$) and 
energy $\epsilon_0=\hbar^2/(2mx_0^2)$ [$V(\xi,\tau)=V_{ext}(x,t)/\epsilon_0$]. 
From the SP wavefunctions $\psi_m$ one first constructs a fully 
antisymmetric (fermionic) wavefunction in the form of the Slater determinant, 
$\psi_F(x_1,\ldots,x_N,t)=\sqrt{x_0^N/N!} \det [\psi_m(\xi_j,\tau)]$; 
$\psi_F$ describes a system of spinless noninteracting fermions in the 
1D potential $V_{ext}(x,t)$ \cite{Girardeau1960,Girardeau2000}. The bosonic 
many-body solution $\psi_B(x_1,\ldots,x_N,t)$ is obtained after symmetrization 
of $\psi_F$:

\begin{equation}
\psi_B=A(x_1,\ldots,x_N)\sqrt{\frac{x_0^N}{N}}
\det_{m,j=1}^{N} [\psi_m(\xi_j,\tau)],
\label{psi_B}
\end{equation} 
where $A=\Pi_{1\leq i < j\leq N} \mbox{sgn}(x_i-x_j)$ is a "unit antisymmetric function" 
\cite{Girardeau1960}. Thus, the quantum dynamics of the TG gas is obtained 
from Eq. (\ref{psi_B}) after solving Eq. (\ref{master}). For example, the 
evolution of the single-particle density 
$\rho_{SP}(x,t)=\int dx_2\ldots dx_N |\psi_B(x,x_2,\ldots,x_N,t)|^2$ corresponds 
to the evolution of $\rho(\xi,\tau)=\sum_m|\psi_m(\xi,\tau)|^2$ \cite{Girardeau2000}. 
It should be noted that while some quantities of the corresponding fermionic 
system are identical (e.g., the SP density), some significantly differ 
(e.g., the momentum distribution \cite{Lenard1964,Paredes2004}).

Excited many-body eigenstates with soliton-like SP density are found in 
real, time-independent, and parity invariant potentials, $V(\xi)=V(-\xi)$. 
Let $\phi_{\epsilon,\gamma}(\xi)$ denote the eigenstates, and let 
$\epsilon$ denote eigenvalues (energies) of the SP Hamiltonian 
$H=-d^2/d\xi^2+V(\xi)$ with appropriate boundary conditions. The 
extra index $\gamma$ is used in case there are degenerate SP eigenstates. 
Since the Hamiltonian commutes with the parity operator, they can have 
a common complete set of eigenstates, in which case the eigenstates 
$\phi_{\epsilon,\gamma}(\xi)$ are either symmetric 
$\phi_{\epsilon,\gamma}^{+}(\xi)=\phi_{\epsilon,\gamma}^{+}(-\xi)$ 
(even parity) or antisymmetric $\phi_{\epsilon,\gamma}^{-}(\xi)=
-\phi_{\epsilon,\gamma}^{-}(-\xi)$ (odd parity). Consider a bosonic 
many-body wavefunction, constructed according to Eq. (\ref{psi_B}), 
from odd-parity eigenmodes only, i.e., every 
$\psi_m$ equals to one of the eigenstates $\phi_{\epsilon,\gamma}^{-}(\xi)$
($\psi_m\neq\psi_n$ for $m\neq n$). 
Such a wavefunction is an {\em excited many-body eigenstate} of the 
system. Its SP density $\rho^{-}(\xi)=\sum |\phi_{\epsilon,\gamma}^{-}(\xi)|^2$ 
is stationary, and $\rho^{-}(0)=0$ because $\phi_{\epsilon,\gamma}^{-}(0)=0$. 
The SP density of this excited many-body eigenstate thus has a dip at $\xi=0$, 
which resembles the structure of nonlinear dark-solitons 
\cite{Dum1998,ExpDark,Busch2000,Muryshev2002}. Therefore we will refer to these states 
as dark many-body eigenstates. If we construct waves $\psi_m$ from symmetric 
eigenmodes $\phi_{\epsilon,\gamma}^{+}(\xi)$ only, the density 
$\rho^{+}(\xi)=\sum |\phi_{\epsilon,\gamma}^{+}(\xi)|^2$ is likely to have a 
pronounced peak at $\xi=0$, reminiscent of anti-dark solitons. 
Hence, they will be referred to as anti-dark many-body eigenstates.

\begin{figure}
\includegraphics[scale=0.34,angle=-90]{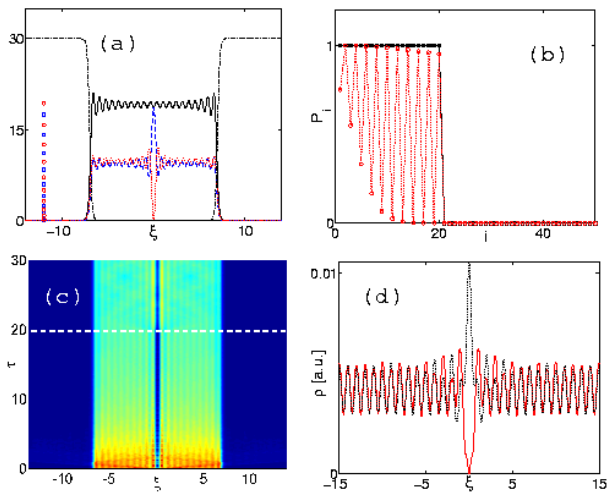}
\caption{
(color online) Dark and anti-dark many-body eigenstates: structure and 
excitation. 
(a) The container potential $V_c(\xi)$ (black dot-dashed line), 
the even $\rho^{+}$ (blue dashed line), odd $\rho^{-}$ (red dotted line), 
and total SP density $\rho=\rho^{+}+\rho^{-}$ (black solid line)
of the ground state with $20$ bosons; 
the first $20$ eigenvalues are shown as blue squares (even SP 
eigenstates) and red circles (odd SP eigenstates).
(b) The spectrum $P_i$ of the excitation at $\tau=0$ and $\tau=19.8$, and  
(c) the dynamics of the SP density following the time-dependent perturbation.
(d) The SP density $\rho(\xi)$ of an anti-dark (black dotted line) and 
dark (red solid line) many-body eigenstate in the periodic potential 
(see text for details).
}
\label{fig1}
\end{figure}

Dark and anti-dark many-body eigenstates are specially tailored 
excitations of the TG gas, constructed solely from the odd or 
even SP eigenstates of the parity invariant potential $V(\xi)$. 
Assuming the absence of degeneracy, the even and odd parity SP 
eigenstates alternate when ordered with respect to energy, meaning 
that such specially tailored states are unlikely to naturally occur. 
We illustrate this by studying $N=20$ TG bosons in the external 
container-like potential 
$V_c(\xi)=V_c^0 \{ 2 + \sum_{i=1,2} (-)^{i+1}\tanh x_w(\xi+(-)^i x_c) \}$ 
($V_c^0=15$, $x_w=4$, and $x_c=7$), shown in Fig. \ref{fig1}(a). 
The SP density 
$\rho$ of the ground state is plotted as a solid black line, whereas 
the dotted red (dashed blue) line depicts the odd (even, respectively) 
component of the SP density: $\rho(\xi)=\rho^{+}(\xi)+\rho^{-}(\xi)$. 
The energies of the even (odd) parity SP eigenmodes are shown as blue 
squares (red circles, respectively). As expected odd and even parity 
eigenstates alternate with increasing energy. Thus, half of the 
SP eigenstates comprising the ground state are even and half are odd. 
The even $\rho^{+}$ (odd $\rho^{-}$) component of the SP density has 
the structure of the anti-dark (dark, respectively) many-body eigenstates. 
In what follows we propose a method for the dynamical excitation 
of dark stationary waves. 

Our scheme utilizes a time-dependent potential which tailors 
the many-body wavefunction in a specific desired fashion, and 
separates the odd from the even component. The system of $N=20$ 
TG bosons is initially in the ground state of the 
confining potential, see Fig \ref{fig1}(a). In the spirit of Refs. 
\cite{Girardeau2000,Busch2003}, we perturb this system at its symmetry 
point $\xi=0$, with a spatially-narrow repulsive potential, which may 
be obtained with a laser \cite{Girardeau2000,Busch2003}. However, in 
contrast to \cite{Girardeau2000,Busch2003}, here we periodically switch 
this potential on and off in time. Such a time-dependent potential can 
be modeled as $V'(\xi,\tau)=V'_0[\mbox{sgn}(\sin(2\pi t/\tau_p))+1)]
\exp[-(\xi/\sigma)^2]$, where $\tau_p=0.1$ is the periodicity of the 
laser signal, $V'_0=100$ corresponds to the peak intensity of the laser, 
and $\sigma=0.06$ to its spatial focusing width. As bosons are kicked 
by the time-dependent potential, they acquire energy which has a certain 
probability of being higher than the lip of the trap and are ejected 
away from it. However, because the laser is focused close to $\xi=0$, 
it strongly affects only the even-parity SP eigenstates, whereas the 
odd-parity eigenstates are left nearly unperturbed. Consequently, the 
many-body wavefunction within the container $(|x|<x_c)$ takes on a 
specific structure: it is constructed via Eq. (\ref{psi_B}) mainly 
from the odd-parity SP eigenstates. This filtering process is 
depicted in Fig. \ref{fig1}(b), showing the spectrum of the SP 
wavefunctions $\psi_m(\xi,\tau)$ calculated according to 
$P_i(\tau)=\sum_m |\int_{-2x_c}^{2x^c} d\xi \psi_m(\xi,\tau)\phi_i^*(\xi)|^2$, 
where $\phi_i(\xi)$ is the $i$th eigenstate of the SP Hamiltonian. 
The spectrum at $\tau=0$ is flat (black squares) because the odd and the 
even eigenstates are equally present. However, the spectrum after 
$\tau=19.8$ (red circles) is mainly comprised from odd-parity eigenstates. 
Fig. \ref{fig1}(c) shows the evolution of the total SP density. The 
time-dependent potential acts within the interval $\tau=[0,19.8]$. 
After $\tau=19.8$ (marked by a horizontal line) it is turned off. 
The SP density nevertheless retains a dark notch at $\xi=0$ even after 
the time-dependent potential is turned off, clearly displaying dark 
stationary wave evolution. The numerical evolution of Eq. (\ref{master})
is performed with the split-step Fourier method.

Although different in nature, the scheme proposed here is akin 
to evaporative cooling. The concept should work for various types 
of container-like potentials. The scheme is also fairly robust. For 
lasers with larger intensity $\propto V'_0$, the filtering occurs 
on faster time scales (i.e., a smaller number of on-off switches 
are sufficient) and is more efficient. The focusing width of the 
laser $\sigma$ limits the number of particles which can be efficiently 
filtered. This width should be sufficiently smaller than the period 
of the spatial oscillation (close to $\xi=0$) of the $N$th SP 
eigenstate. It should be emphasized that even though we follow 
the spirit of Refs. \cite{Girardeau2000,Busch2003}, the proposed 
scheme separates the odd and the even component 
in space, while the many-body wavefunction within the container 
assumes the particular structure of a dark stationary wave.

While the proposed method for exciting dark stationary states 
of the TG gas employs a container-like potential, it should be 
emphasized that the notion of dark and anti-dark many-body 
eigenstates pertains to various parity invariant potentials. 
We illustrate this fact in a periodic potential $V(\xi)=V(\xi+D)$ 
(e.g., optically induced lattices). The SP eigenstates of this 
system are Bloch waves \cite{E-M} of the form 
$\phi_{k,n}(\xi,\tau)=u_{k,n}(\xi) e^{ik\xi}e^{-i\epsilon_{k,n} \tau}$, 
where $n$ denotes the band number, $k$ is the Bloch wave-vector, 
and $u_{k,n}(\xi)=u_{k,n}(\xi+D)$ describes the periodic spatial 
profile of the Bloch wave. Since $V(\xi)=V(-\xi)$, the Bloch 
waves $\phi_{k,n}$ and $\phi_{-k,n}$ are degenerate. By properly 
choosing the coefficients $a_{k_m}$ and $a_{-k_m}$ within the 
superposition $\psi_{m}(\xi,\tau)=a_{k_m}\phi_{k_m,n}+ a_{-k_m} \phi_{-k_m,n}$, 
degenerate eigenstates $\phi_{\pm k_m,n}$ can be superimposed 
to obtain even [$\psi_m^{+}(\xi,\tau) = \psi_m^{+}(-\xi,\tau)$] 
and odd-parity [$\psi_m^{-}(\xi,\tau) = -\psi_m^{-}(-\xi,\tau)$] 
eigenstates. The many-body wavefunction comprised solely from 
$\psi_m^{-}(\xi,\tau)$ [$\psi_m^{+}(\xi,\tau)$] via Eq. (\ref{psi_B}) 
is a dark (anti-dark) excited many-body eigenstate of the TG gas 
in the lattice. Figure \ref{fig1}(d) shows the (stationary) SP 
density of the dark and anti-dark many-body eigenstate in a periodic 
potential $V(\xi)=10 \cos^2(\pi x)$, constructed by symmetrizing 
the lowest $N=21$ SP eigenmodes. The calculation is performed on 
the ring (periodic boundary conditions) of length $L=71$.

{\em Relation to linear incoherent light.-} Dark and anti-dark 
excited many-body eigenstates of the TG gas are analogous to 
partially coherent nondiffracting beams that were studied in 
the context of classical optics \cite{Turunen,Shchegrov}. 
In order clarify this point, we first demonstrate the mapping 
between the propagation of incoherent light in linear 1D photonic 
structures \cite{Joan} and the TG gas dynamics. Consider a 
quasimonochromatic, linearly polarized, partially-spatially 
incoherent light beam which propagates paraxially in the 1D 
photonic structure described by the spatially dependent index 
of refraction $n_{tot}^2=n_0^2+2 n_0 n(x,z)$. The classical 
electromagnetic field $E(x,z,t)$ of the beam randomly fluctuates; 
$z$ ($x$) denotes the propagation axis (spatial, respectively) 
coordinate. The state of the system is described by the mutual 
coherence function $B=\langle E^*(x_2,z,t)E(x_1,z,t) \rangle$ \cite{Wolf}, 
where brackets denote the time-average, which equals ensemble average 
assuming the light source is stationary and ergodic \cite{Wolf}. 
The mutual coherence function $B$ can be decomposed through an 
orthonormal set of coherent modes $\tilde\psi_m$ and their modal 
weights $\lambda_m$ \cite{Wolf},

\begin{equation}
B(x_1,x_2,\tau)\sum_m \lambda_m \tilde\psi_m^*(x_2,z) \tilde\psi_m(x_1,z).
\label{B}
\end{equation}
In order to connect $B$ to Eq. (\ref{master}) we switch to 
dimensionless units: $\xi=x/x_0$, $\tau=z/(2kx_0^2)$ 
("time" is here the propagation length) where $k=n_0\omega/c$, 
and $\omega$ is the temporal frequency of the beam. The potential 
$V$ arises from the refractive index $V(\xi,\tau)=-2(kx_0)^2 n(x,z)/n_0$. 
Waves $\psi_m(\xi,\tau)=\sqrt{ x_0}\tilde\psi_m(x,z)$ obey Eq. 
(\ref{master}) (e.g., see Ref. \cite{Equ}), affirming the mapping between 
the two systems. Note that each solution of Eq. (\ref{master}) 
generates one bosonic many-body wavefunction via Eq. (\ref{psi_B}),
and, due to the arbitrary choice of modal weights $\lambda_m$, 
many correlation functions (\ref{B}) corresponding to 
incoherent optical fields propagating in linear 1D photonic structures. 
The density $\rho=\sum_m|\psi_m|^2$ corresponds to 
time-averaged intensity $I=\sum_m\lambda_m |\tilde\psi_m|^2$ \cite{Wolf}.

One particular example of a partially coherent nondiffracting 
optical beam propagating in vacuum, corresponds to the dark 
stationary TG wave on a ring studied by Girardeau and Wright 
\cite{Girardeau2000}. An incoherent optical beam with the mutual 
coherence function $B(x_1,x_2)=\int dk_x G(k_x) \sin(k_x x)$, or 
equivalently with the modal structure $\tilde\psi_{k_x}(x)=\sqrt{G(k_x)} \sin(k_x x)$, 
is propagation invariant. If its power spectrum $G(k_x)$ is rectangular, 
$G(k_x)=I_0/K$ for $|k_x|<K$, and zero otherwise, the intensity 
structure has the form $I_0/2[1 - j_0(Kx)]$, which is {\em exactly} 
the form of the odd SP density-component of the dark stationary wave 
on a ring in the thermodynamic limit \cite{Girardeau2000}. If 
$B(x_1,x_2)=\int dk_x G(k_x) \cos(k_x x)$, one obtains anti-dark 
optical propagation-invariant waves.

We have thus established a link between the dynamics of incoherent 
light in linear photonic media and TG gas via Eq. (\ref{master}). It 
should be kept in mind the the former system is {\em classical}, while 
the latter is {\em quantum}, and the evolution of the quantities derived 
from the set of waves $\psi_m$ (e.g., the density $\rho=\sum_m|\psi_m|^2$) 
should be properly interpreted. This mapping adds to the analogies between 
optical and matter waves \cite{NAO}, and in particular to the analogy 
between nonlinear partially coherent optical- and matter-waves \cite{Buljan2005}.

Before closing, it should also be noted that the evolution equation 
for the mutual coherence function $B(x_1,x_2,z)$, in the paraxial 
approximation, in linear photonic structures (e.g., see \cite{Equ}), 
is identical to the evolution of the reduced single-particle density 
matrix of non-interacting spinless fermions (in 1D, and 2D as well). 

In conclusion, we have proposed a scheme for exciting dark stationary 
waves of the TG gas. Within our scheme, a time-dependent potential 
focused to the center of the trap, selectively "evaporates" a non-desirable 
part of the many-body wavefunction, thereby creating a dark stationary wave. 
The stationary waves of the TG gas are analogous to partially-coherent 
nondiffracting beams in optics. This analogy is a consequence of the 
mapping between incoherent light in linear 1D photonic structures 
and the TG gas.


\begin{references}

\bibitem{Girardeau1960}
M. Girardeau,
J. Math. Phys. {\bf 1}, 516 (1960). 

\bibitem{Lieb}
E. Lieb and W. Lineger, 
Phys. Rev. {\bf 130}, 1605 (1963);
E. Lieb, Phys. Rev. {\bf 130}, 1616 (1963).

\bibitem{Lenard1964}
A. Lenard, J. Math. Phys. {\bf 5}, 930 (1964). 

\bibitem{Girardeau2000}
M. Girardeau and E.M. Wright,
Phys. Rev. Lett. {\bf 84}, 5691 (2000). 

\bibitem{OneD}
F. Schrek {\em et al.}, 
Phys. Rev. Lett. {\bf 87}, 080403 (2001);
A. G\" orlitz {\em et al.}, 
Phys. Rev. Lett. {\bf 87}, 130402 (2001); 
M. Greiner {\em et al.}, 
Phys. Rev. Lett. {\bf 87}, 160405 (2001);
H. Moritz {\em et al.}, 
Phys. Rev. Lett. {\bf 91}, 250402 (2003);
B.L. Tolra {\em et al.},
Phys. Rev. Lett. {\bf 92}, 190401 (2004);
T. St\" oferle {\em et al.},
Phys. Rev. Lett. {\bf 92}, 130403 (2004). 

\bibitem{Kinoshita2004}
T. Kinoshita, T. Wenger, and D.S. Weiss,
Science {\bf 305}, 1125 (2004). 

\bibitem{Paredes2004}
B. Paredes {\em et al.},
Nature (London) {\bf 429}, 377 (2004). 

\bibitem{Kinoshita2006}
T. Kinoshita, T. Wenger, and D.S. Weiss,
Nature (London) {\bf 440}, 900 (2006). 

\bibitem{Olshanii}
M. Olshanii, Phys. Rev. Lett. {\bf 81}, 938 (1998).

\bibitem{Petrov}
D.S. Petrov, G. Schlyapnikov, and J.T.M. Valraven, 
Phys. Rev. Lett. {\bf 85} 3745 (2000). 

\bibitem{Dunjko}
V. Dunjko, V. Lorent, and M. Olshanii,
Phys. Rev. Lett. {\bf 86} 5413 (2001). 

\bibitem{Rigol2006}
M. Rigol {\em et al.},
arXiv:cond-mat/0604476 (2006). 

\bibitem{Busch2003}
T. Busch and G. Huyet, 
J. Phys. B {\bf 36} 2553 (2003). 

\bibitem{Girardeau2000a}
M. Girardeau and E.M. Wright,
Phys. Rev. Lett. {\bf 84}, 5239 (2000). 

\bibitem{Ohberg2002}
P. Ohberg and L. Santos, 
Phys. Rev. Lett. {\bf 89}, 240402 (2002).

\bibitem{Rigol2005}
M. Rigol and A. Muramatsu,
Phys. Rev. Lett. {\bf 94}, 240403 (2005).

\bibitem{Berman2004}
G.P. Berman {\em et al.},
Phys. Rev. Lett. {\bf 92}, 030404 (2004).

\bibitem{Minguzzi2005}
A. Minguzzi and D.M. Gangardt, 
Phys. Rev. Lett. {\bf 94}, 240404 (2005).

\bibitem{Dum1998}
R. Dum {\em et al.},
Phys. Rev. Lett. {\bf 80}, 2972 (1998).

\bibitem{ExpDark}
S. Burger {\em et al.},
Phys. Rev. Lett. {\bf 83}, 5198 (1999);
J. Denschlag, {\em et al.},
Science {\bf 287}, 97 (2000).

\bibitem{Busch2000}
Th. Busch, and J.R. Anglin,
Phys. Rev. Lett. {\bf 84}, 2298 (2000).

\bibitem{Muryshev2002}
A. Muryshev {\em et al.},
Phys. Rev. Lett. {\bf 89}, 110401 (2002).

\bibitem{Turunen}
J. Turunen, A. Vasara, and A.T. Friberg, 
J. Opt. Soc. Am. A {\bf 8} 282 (1991). 

\bibitem{Shchegrov}
A.V. Shchegrov and E. Wolf,
Opt. Lett. {\bf 25} (2000).

\bibitem{E-M}
N.W. Aschroft and N.D. Mermin,
{\em Solid State Physics}
(Saunders, Philadelphia, 1976).

\bibitem{Joan}
J.D. Joannopoluos, R.D. Meade, J.D. Winn,
{\em Photonic Crystals: Molding the Flow of Light},
(Princeton University Press, Princeton, 1995). 

\bibitem{Wolf}
L. Mandel and E. Wolf, 
{\em Optical Coherence and Quantum Optics},
(Cambridge University Press, New York, 1995). 

\bibitem{Equ}
D.N. Christodoulides {\em et al.}, 
Phys. Rev. E {\bf 63}, 035601 (2001).

\bibitem{NAO}
G. Lens, P. Meystre, and E.M. Wright,
Phys. Rev. Lett. {\bf 71}, 3271 (1993);
S. L. Rolston and W. D. Phillips, Nature {\bf 416}, 219 (2002).

\bibitem{Buljan2005}
H. Buljan, M. Segev, and A. Vardi,
Phys. Rev. Lett. {\bf 95}, 180401 (2005). 


\end{references}
\end{document}